\documentclass[twocolumn,secnumarabic,amssymb, nobibnotes, aps, pre]{revtex4-2}  

\usepackage{graphicx}
\usepackage{chemformula}

\usepackage{hyperref}  

\usepackage{xcolor}

\newcommand{\ie}{\textit{i.e.}}

\newcommand{\etal}{\textit{et al.}}

\newcommand{\um}{~\(\mu\text{m}\)}
\newcommand{\degC}{\(^\circ\)C}

\begin{document}

\title{Pattern of inclusions inside rippled icicles}

\author{John Ladan}
\email{jladan@physics.utoronto.ca}
\affiliation{Department of Physics, University of Toronto, \\ 60 St. George St., Toronto, ON Canada M5S 1A7}
\author{Stephen W. Morris}
\email{smorris@physics.utoronto.ca}
\affiliation{Department of Physics, University of Toronto, \\ 60 St. George St., Toronto, ON Canada M5S 1A7}

\begin{abstract}
    Icicles that have grown from slightly impure water develop ripples around
    their circumference.  The ripples have a near-universal wavelength and are
    thought to be the result of a morphological instability. Using
    laboratory-grown icicles and various species of impurities, including
    fluorescent dye, we show that a certain fraction of the impurities remain
    trapped inside the icicle, forming inclusions within the ice.  The
    inclusions are organized into chevron patterns aligned with the peaks of the
    ripples.  Within the chevrons, a substructure of crescent-shaped structures
    is observed.  We also examine the crystal grain structure of laboratory
    icicles, with and without impurities.  We present the first detailed study
    of these growth patterns in the interior of icicles, and discuss their
    implications for the mechanism of the ripple-forming instability.
\end{abstract}

\maketitle

\section{Introduction}
\label{sec:introduction}

\begin{figure}
 \includegraphics[width=2.3in]{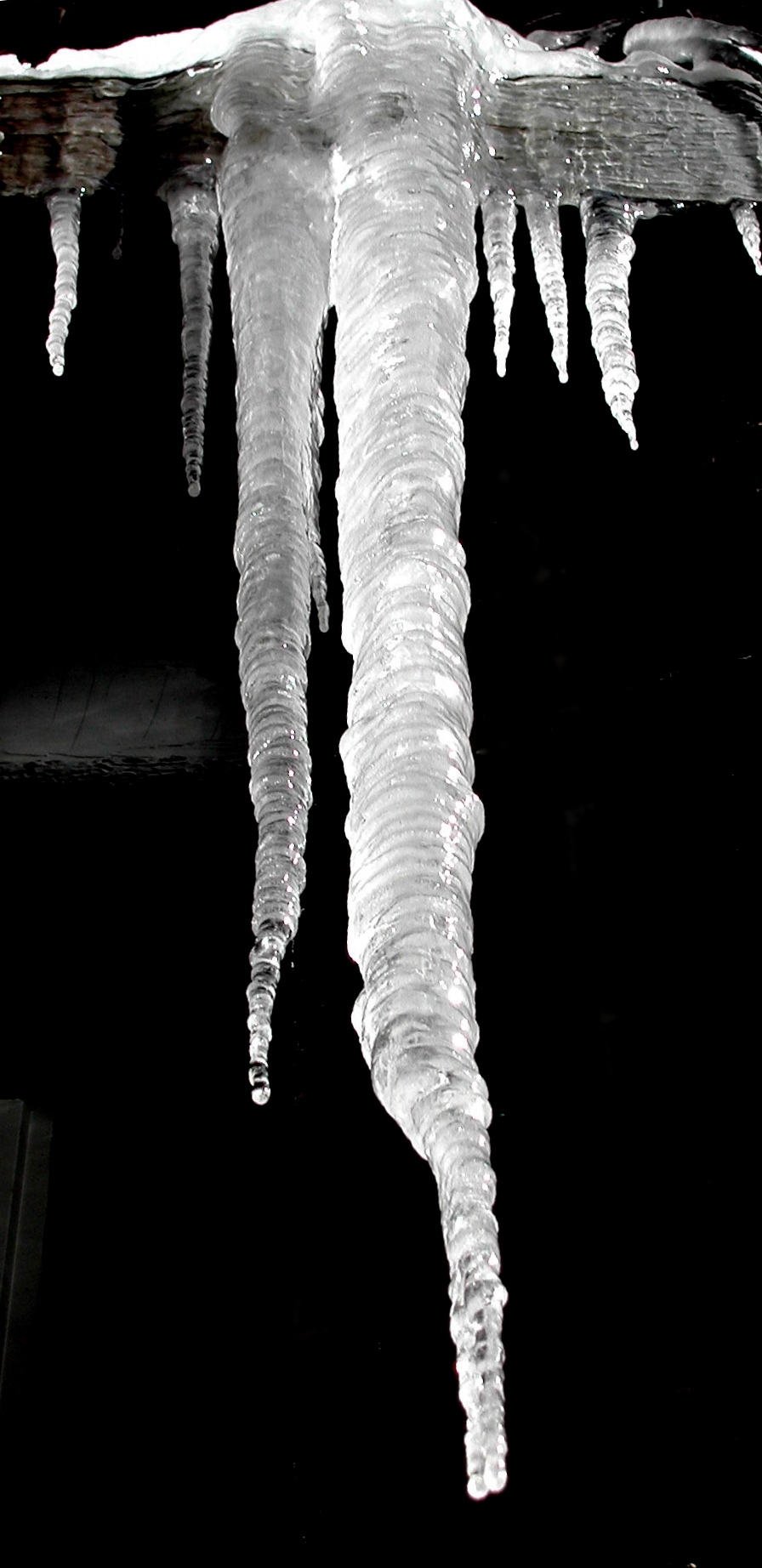}
 \caption{An upward looking view of ripply natural icicles hanging from the edge of a roof.}
  \label{fig:roof-icicles}
\end{figure}

Icicles have a distinctive shape which is the result of a complex growth
process~\cite{maeno1984a,maeno1984b,maeno1994}.  Water flowing from a point of
support into sub-freezing air forms an elongated ice structure which grows both
in length and, more slowly, in width, with excess water dripping from the
growing tip~\cite{makkonen1988Model}.  Small amounts of impurities in the feed
water have profound effects on the evolving shape, triggering characteristic
ripples that emerge around the circumference~\cite{chen2013njp}, as well as
having other effects on the overall shape~\cite{chen2011pre}.   
{While dissolved impurities have been shown experimentally
to lead to ripples,
the underlying mechanism of this instability
remains an open problem.}
Impure icicles
also exhibit a foggy appearance, compared to ones grown from pure water. 
While
atmospheric precipitation is typically quite pure, many naturally occurring
icicles on structures nevertheless exhibit clear effects of impurities, as shown
in Fig.~\ref{fig:roof-icicles}.

In this paper, we build on previous laboratory work which uncovered the role of
impurities in ripple formation~\cite{chen2013njp}, and which also revealed the
complex nature of the exterior flow over the icicle
surface~\cite{ladan2021wetting}.  Here, we study the distribution of trapped
impurities and the crystal structure in the interior of the icicle, and how they
are related to the exterior rippled shape.  Our experiments are intended to shed new light on the elusive mechanism of the 
impurity-driven ripple formation.

Our work extends earlier
observations of natural icicles by Laudise and Brand~\cite{laudise1979} and
Knight~\cite{knight1980}, and the laboratory work of Maeno \etal{}~\cite{maeno1984b,maeno1994}.  Using controlled icicle growth and dye
techniques, we present the first detailed study of the pattern of impurities
that remain in the ice and their relationship to the topography of the ice
surface.  We also studied the crystal structure and its relationship to the
ripples and impurities. Our apparatus and Methods are described in
Sec.~\ref{sec:experiment} below.

The polycrystalline nature of icicles is well known~\cite{walker1988SciAm}.  It
has also long been observed that some icicles contain significant amounts of
unfrozen liquid water within their ice matrix --- so-called \textit{spongy
ice}~\cite{walker1988SciAm,knight1980,maeno1984b,maeno1994}.
Knight~\cite{knight1980} attributed the foggy appearance of rippled icicles to
\textit{air bubbles} trapped inside the ice, which he proposed was the result of
more rapid cooling near the ripple peaks. Knight further suggested that the
sponginess indicated a particular crystal orientation, the c-axis parallel to
icicle axis, which conflicted with the observations of Laudise and
Brand~\cite{laudise1979}, who found that the c-axis was mainly perpendicular to
the icicle axis. These reports on natural icicles were made before the strong
connection between feed-water impurities and ripples was
established~\cite{chen2013njp}.

In this paper, we show in Sec.~\ref{sec:observations} that the foggy appearance
is not due to trapped air bubbles, but rather due to small features we call
\textit{inclusions} which are actually pockets of highly impure liquid trapped
inside the ice matrix. We examine the spatial distribution of the inclusions,
which are organized into bands making a \textit{chevron} pattern.  Within the
chevrons, we find that the inclusions are further organized into smaller
substructures we call \textit{crescents}.  Chevron patterns in natural icicles
were previously noted by Maeno \etal{}~\cite{maeno1984b,maeno1994}. The chevrons
and crescents align closely with the growth and upward
migration~\cite{chen2013njp} of ripples on the icicle surface, as discussed in
Sec. ~\ref{sec:connection-to-ripples}. In Sec.~\ref{sec:grain-structure}, we
visualize the crystal grain structure within the ice and show that it is
sometimes correlated with the ripple growth, with the liquid inclusions trapped
within the crystallites, not at their boundaries. 

In Sec.~\ref{sec:discussion}, we discuss how the pattern of inclusions may be
related to the patchy nature of the surface liquid
coverage~\cite{ladan2021wetting}, and how our observations could inform future
models of icicle growth.  Sec.~\ref{sec:conclusions} summarizes our conclusions
and the remaining open questions.

\section{Experiment}
\label{sec:experiment}

To make icicles in the laboratory, we used a purpose-built icicle growing
machine, which has been described in detail in several previous
publications~\cite{chen2011pre,chen2013njp,ladan2021wetting}.  Briefly, it
consisted of a refrigerated box filled with temperature controlled, turbulent
air.  The humidity in the box was continuously measured and was typically 85\%
during active icicle growth.   
Feed water was introduced at the top of the box at a controlled temperature and
flow rate, and fell onto a slowly rotating conical support.  The support
rotation period of 8 minutes allowed all sides of the growing icicle to be
visualized from the side of the box by a 36 MP SLR camera (Nikon D810).  The
rotation was indexed, so that images were acquired at 16 equally spaced
rotational positions.  This allowed the complete 3D shape of the icicle to be
reconstructed using edge detection throughout its growth.  To measure the
inclusions inside the icicle, it was removed from the box for further analysis,
as described in Sec.~\ref{sec:methods} below.

A total of 51 icicles were grown, using distilled water with NaCl (30 icicles),
Dextrose (5 icicles), or Sodium Fluorescein (16 icicle) as an impurity. The
icicles were grown in several series, starting with the highest impurity
concentration, followed by feed water diluted by half for each subsequent
icicle. 
For one quarter of the icicles, the feed water was degassed by under vacuum
while stirred by a magnetic stirrer for 10 minutes directly before the icicle
machine was started. In our tests, this method reduced dissolved oxygen by
$\sim 95\%$, with less than 0.01\% loss of water.
The typical feed-water flow rate was 3.0~g/min, apart from one series of
three Sodium Fluorescein icicles. The air temperature was held at values between
-14.0\degC{} and -11.5\degC{}, consistent within $\pm0.25$\degC{} for each
series (-11.5\degC{}, -12.5\degC{}, -13.5\degC{} for Dextrose, and
-14.0\degC{}).

{The ionic concentration of source water used ranged between 0.04~mMol/kg
to 43.8~mMol/kg, which is equivalent to between 1.2~ppm~NaCl to 1280~ppm~NaCl.
This covers a wider range than would be expected for natural icicles. 
Chen \etal{} measured a typical concentration of 0.25~mMol/kg in snow from a rooftop and in melted natural icicles~\cite{chen2013njp}. This was compared to 2.9~mMol/kg for tap water from Toronto Ontario, which forms icicles with prominent ripples.  Melted natural icicles and their source water are generally intermediate in purity between distilled water and Toronto tap water.

\subsection{Measuring the topography and inclusions}
\label{sec:methods}



The topography of each icicle was extracted from the time-lapse photography
during growth using a canny edge detector, then refined using a simple
``in-painting'' algorithm. Peaks in the topography were found with the Python
function \verb;scipy.signal.find_peaks();~\cite{2020SciPy-NMeth} with
\verb;prominence=3; (3 pixels is approximately 0.2 mm). Several examples of the
mapped topography and the extracted peaks are shown in
Fig.~\ref{fig:space-time-peaks}.  Each icicle was imaged over 700 times during
its growth, resulting in detected edges consisting of tens of thousands of data
points. We thus have rather complete information about the evolution of the
icicle's external topography during growth, which can then be compared to the
distribution of impurity inclusions at the end of growth process, which could be
obtained by sectioning the icicle at a known rotational position.  


After growing, the icicle and its support were removed from the box, and the
icicle was sawed off below the tip of the support cone.  It was then stored in a
freezer at -24\degC{} until sectioning.
Cross sections were taken parallel to the icicle axis by melting each icicle on
aluminum plate to approximately 3mm thickness. 
{The section thickness was
selected to ensure there were enough interior inclusions visible to see a clear
pattern, without occlusions from off-center layers.} 

Images of the sections were taken in a chilled box with a unidirectional light
source, which scattered off inclusions inside the ice. The sections were placed
on a black rubber mat with a steel ruler placed beside so that the resolution of
each image could be measured. 
{Before imaging, the sections were wiped
with an absorbent
paper towel to remove liquid on the surface and inside exposed inclusions.} For the fluorescent dye, the icicles were
illuminated with a strip of UV LEDs. The camera (a Canon EOS Rebel T2i SLR with
a Canon EF-S 18-55mm lens) was positioned above the chilled box on a frame, and
the images were later measured using ImageJ~\cite{schindelin2012Fiji}.  The
results of this analysis are presented in Sec.~\ref{sec:connection-to-ripples}
below.

\begin{figure*}
    \begin{center}
        \includegraphics[width=\textwidth]{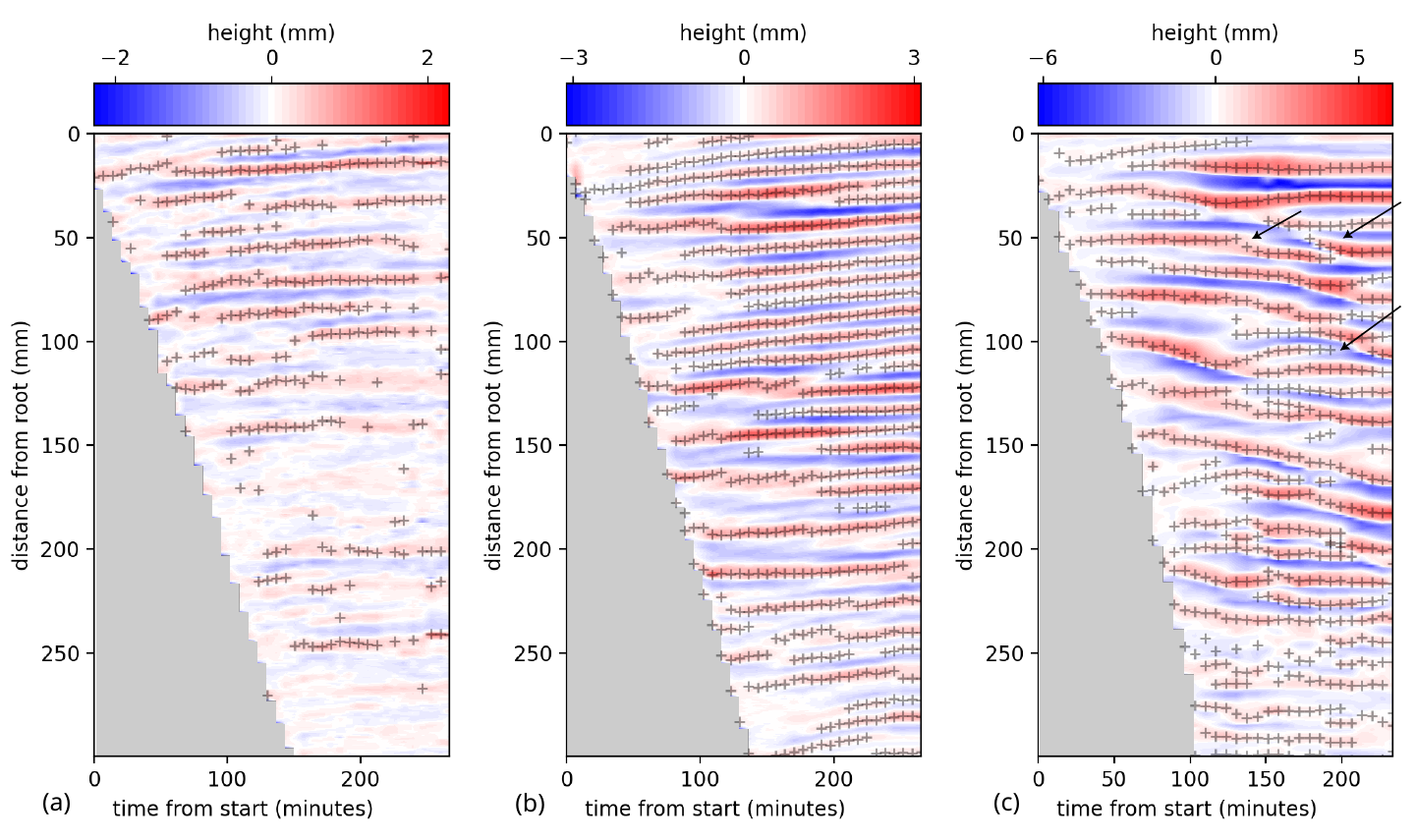}
    \end{center}
    \caption{The extracted topography of icicles with concentrations
    (a) 0.34~mMol/kg (10~ppm~NaCl), (b) 2.7~mMol/kg (80~ppm~NaCl),
    (c) 22~mMol/kg (640~ppm~NaCl). The peaks of ripples are marked by + symbols.
    Note the different color scales. The arrows in part (c) indicate ripple
    splitting or stopping events. }
    \label{fig:space-time-peaks}
\end{figure*}

\section{Observations}
\label{sec:observations}

Inclusions are seen at a wide range of concentrations in icicles grown with
NaCl, as shown in Fig.~\ref{fig:nacl-range}. As the impurity concentration in
the feed water increases, the number of inclusions increases, as does the
amplitude of the the ripples. At the lowest concentrations, the majority of the
icicle is a pure phase of ice (Ice Ih according to the freezing conditions). In
these nearly pure icicles, the inclusions organize themselves into
\textit{crescent} shapes.  These structures within the ice trace the shape of
the ripple peaks and are faintly visible in the upper part of
Fig.~\ref{fig:nacl-range}(a) and more clearly in the microscopic view shown in
Fig.~\ref{fig:microscopic-view}.  The crescents are formed of elongated clusters
of individual inclusions.

As the concentration increases, the crescents of inclusions layer radially upon
one another forming visible \textit{chevron bands} of mixed-phase ice separated
by lines of pure ice. These are most clearly seen in
Fig.~\ref{fig:nacl-range}(c,d,e).  The chevrons are angled upward, reflecting
the upward migration of the ripples during growth at these concentrations. At
320~ppm~NaCl, the single-phase ice is no longer present, and the icicles are
saturated with inclusions, as shown in Fig.~\ref{fig:nacl-range}(g). At these
concentrations, crescents and chevrons are no longer distinguishable from the
general fogginess of the ice.

\begin{figure*}
    \includegraphics[width=\textwidth]{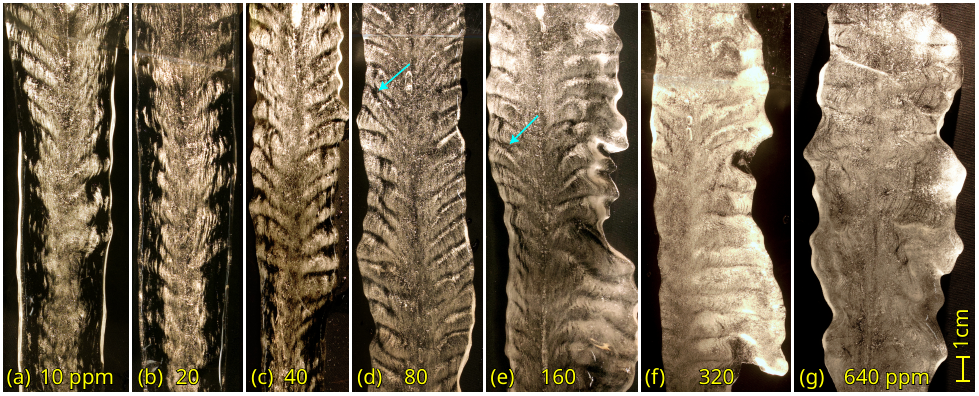}
    \caption{Inclusions are visible at a wide range of concentrations in NaCl
    icicles. At the lowest concentrations (a,b), the majority of the icicle is
    pure ice. As the concentration increases, trapped impurities collect in
    bands that  exhibit a chevron pattern (c,d,e). The transitional concentration
    (f) has thin lines of almost pure ice. At high concentrations (g), the whole
    of the ice is foggy with inclusions. The blue arrows mark ripple stopping events. }
    \label{fig:nacl-range}
\end{figure*}

The small spherical inclusions, which were previously thought to be \textit{air
bubbles}~\cite{maeno1984b,knight1980}, are actually pockets of liquid with a
high concentration of impurities, as we will establish using dye in
Sec.~\ref{sec:nature-of-inclusions}. Examining the inclusions in more detail, as
in Fig.~\ref{fig:microscopic-view}, we find that the inclusions range in
diameter between 20\um{} and 180\um{}. The crescent features are typically
120\um{} in width, and are still present even at higher concentrations when the
chevron bands form.

The inclusions most likely have a spherical shape because their contents never
freeze completely due to the icicle being above the eutectic temperature
(-21.1\degC{} for \ch{H2O + NaCl}).
They will continue to evolve into a spherical shape in equilibrium while the
icicle freezes, and during the rise in temperature as
the section is made by melting on the aluminum plate. Air bubbles would not
undergo the same evolution.

\begin{figure}
    \includegraphics[width=\columnwidth]{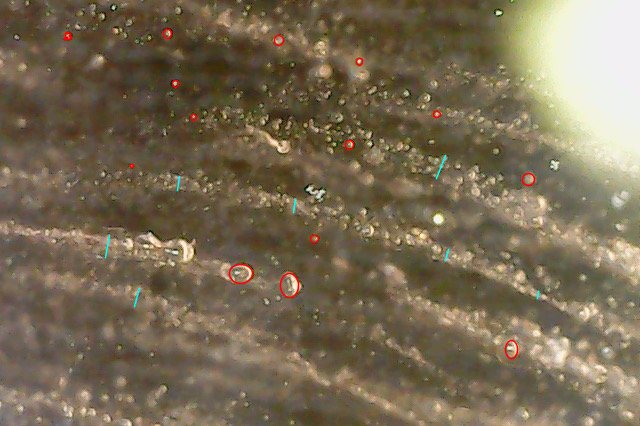}
    \caption{A measurement of the inclusions observed in a section of a
    10~ppm~NaCl icicle near the root. Downward is to the left. The inclusions
    vary significantly in size
    between 20\um{} and 180\um{} diameter. The crescents features have a typical
    width of 120\um{}.}
    \label{fig:microscopic-view}
\end{figure}


If the icicle is cross-sectioned perpendicular to its axis, as shown
in Fig.~\ref{fig:fluorescent-comparison}(c), the crescent
formations of inclusions appear as rings tangential to the icicle surface,
rather like the growth rings of a tree.  These growth rings were observed by
Knight~\cite{knight1980}. 

The shapes of the crescents and chevrons in the longitudinal cross-sections
clearly record important features of the growth process leading to the rippled
exterior shape of the icicle, including the upward migration of the ripples. 
We analyze the relationship between the pattern of inclusions and the shape in
detail in Sec.~\ref{sec:connection-to-ripples}.  The approximate periodicity of
the crescents, in particular, suggests that they are due to cyclic episodes of
wetting and freezing, as we discuss in Sec.~\ref{sec:discussion}.

\subsection{Content of inclusions}
\label{sec:nature-of-inclusions}

Using sodium fluorescein as a fluorescent impurity, we find that the dye is
concentrated into the inclusions, as shown in
Fig.~\ref{fig:fluorescent-comparison}. The solid ice matrix is clear with no dye
present. From this observation, we infer that the impurities are concentrated
inside the inclusions. Because the icicle temperature is never below the
eutectic temperature, the inclusions remain liquid.

Comparing the pattern of inclusions between fluorescein, \ch{NaCl} and glucose
at similar molal concentrations, we find that the inclusions are of similar
size, shape and location. We conclude that the fogginess seen in rippled icicles
is due primarily to scattering by inclusions --- small pockets of liquid with
higher concentrations of impurities --- and not due to air bubbles. We
conjecture that this will be true above the eutectic point for all species of
dissolved impurities that cause constitutional undercooling.  


We did not observe any air bubbles in any of our laboratory icicles. Air bubbles
could only be discerned when a dye is used as the instability-triggering
impurity. When fluorescein is used, there are so many dyed inclusions, that
small non-fluorescent bubbles can not readily be seen.
{When the directional light source is added to the UV illumination,
some inclusions exposed on the surface appear, but this is more likely a result
from removal of liquid when the section is wiped than evidence of air bubbles.}
Maeno \etal{}\ observed oblong inclusions in the core of the
icicle~\cite{maeno1984a,maeno1994}, which could be air bubbles that emerge from
the trapped liquid core (mode 3 growth in the nomenclature of Maeno
\etal{}~\cite{maeno1994}). 
In our fluorescent dyed icicles,  we observe a line of fluorescent inclusions
along the central axis of the icicle where the liquid core freezes last. We
discuss some possible differences between dissolved solids and gases in
Sec.~\ref{sec:discussion}.

\begin{figure}
    \includegraphics[width=\columnwidth]{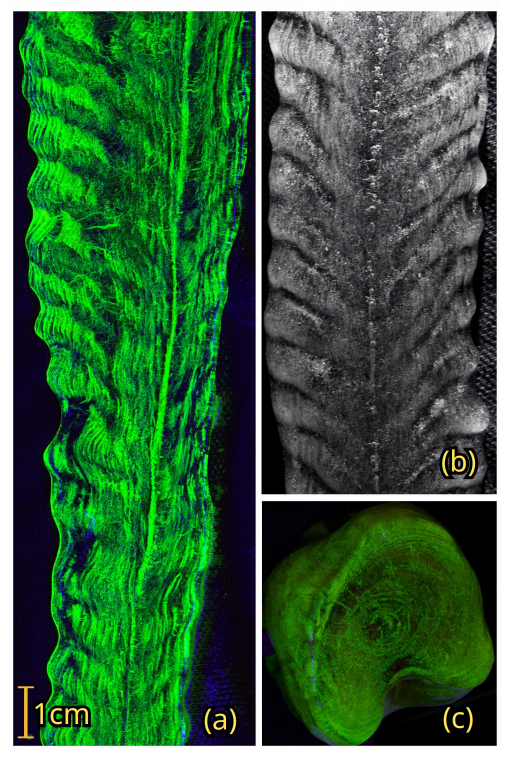}
    \caption{Using Sodium Fluorescein (a) at the same ionic concentration as
    used previously for NaCl, we observe that the inclusions glow under UV
    illumination. The pattern of inclusions qualitatively matches what is
    observed with (b) NaCl (and other chemical species). 
    Some larger inclusions appear near the central line in (b) that may look
    like air bubbles, but the corresponding central line in (a) glows with dye.
    A cross-section perpendicular to the icicle axis (c) shows layers of
    inclusions that resemble the growth rings of a tree.}
    \label{fig:fluorescent-comparison}
\end{figure}

\subsection{\ch{Na+} concentration in the ice}
\label{sec:icp-oes}

The presence of impurities inside of icicles grown with \ch{NaCl} was also
confirmed using ion chromatography by optical emission spectroscopy (ICP-OES).
The feed water, runoff and bulk in-ice concentration were measured for a series
of icicles with ICP-OES~\footnote{Thermo Scientific iCAP Pro ICP OES at the 
Analytical Laboratory for Environmental Science Research and Training (ANALEST),
Department of Chemistry, University of Toronto.}. 
The measured concentrations of \ch{Na+} are presented in
Fig.~\ref{fig:measured-concentrations}.

The bulk concentration of Na+ that we observe is much higher than could be
contained in single-phase ice. The segregation coefficient, $k_0$, which quantifies how
much NaCl can be contained in ice at a given temperature, is essentially zero in
ice Ih, meaning very little NaCl or other impurities can be contained in the
solid ice. Instead, almost all of the impurities are sequestered into liquid
inclusions. The ratio of bulk ice (including inclusions) and feed water
concentrations was found to be nearly constant at $k = 0.27 \pm 0.05$ over the
whole range of concentrations, with only a very slight increasing trend.
%
%

The trapping of NaCl in polycrystaline ice formed by freezing at a constant, nonzero speed has been studied experimentally by Weeks and Lofgren~\cite{weeks1967Effective}, in the context of sea ice formation. They found a solute distribution coefficient $k$, analogous to our bulk ice to feed water concentration ratio, which was growth velocity dependent.  Extrapolating this coefficient to zero growth velocity resulted in an \emph{effective} solute distribution coefficient $k^* = 0.26$ which is remarkably close to our value of this ratio. The value $k^*$ is the effective value of $k_0$ if the ice were to remain in a mixed-phase for a slowly advancing ice front. In their experiment, Weeks and Lofgren used steady growth conditions from a bath of constant concentration, and much higher concentrations from 1\% to 3.3\% NaCl.  }


We did not find a saturation of the bulk ice concentration at high
concentration, as might have been expected. Instead, the bulk ice concentration
increases linearly, so the amount of liquid contained in the ice matrix must
also increase linearly with feed water concentration, even as the icicle becomes
crowded with liquid inclusions. It is possible that there is a saturation point
above the highest measured concentration of 640~ppm~\ch{NaCl}.  Sea water, for
example, is known to form icicles~\cite{chung1990}, but has a far higher NaCl
concentration of about 35~000~ppm.

\begin{figure}
    \begin{center}
        \includegraphics[width=\columnwidth]{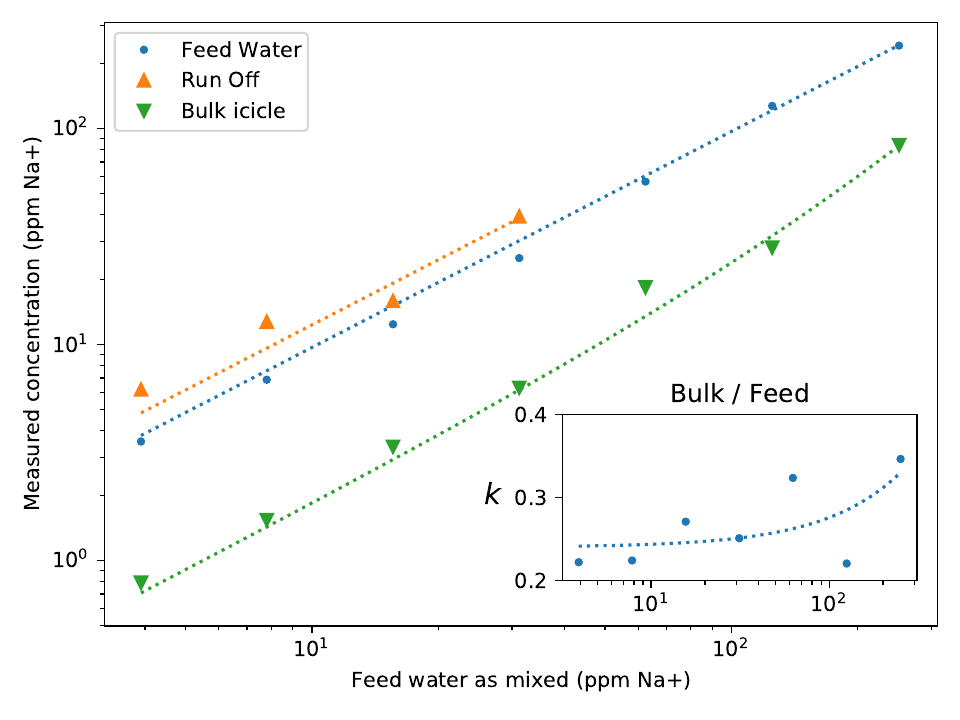}
    \end{center}
    \caption{\ch{Na+} concentrations in the feed water, in runoff water, and in
    cross sections for a series of NaCl icicles, measured with ICP-OES. The feed
    water concentration agrees with what is expected from the amount of NaCl
    added. The runoff water shows only a slightly increased concentration over
    the feed water.  The ratio of concentration in the bulk to that of the feed
    water, $k$,  is nearly constant at $0.27 \pm 0.05$ , but may increase slightly at
    higher concentrations. The inset shows this ratio.  The dotted line in the
    inset is the fit  $k = 0.05 \log_{10} (c) + 0.19$.}
    \label{fig:measured-concentrations}
\end{figure}

\subsection{Connection between inclusions\\and rippled topography}
\label{sec:connection-to-ripples}

\begin{figure}
    \begin{center}
        \includegraphics[height=.5\textheight]{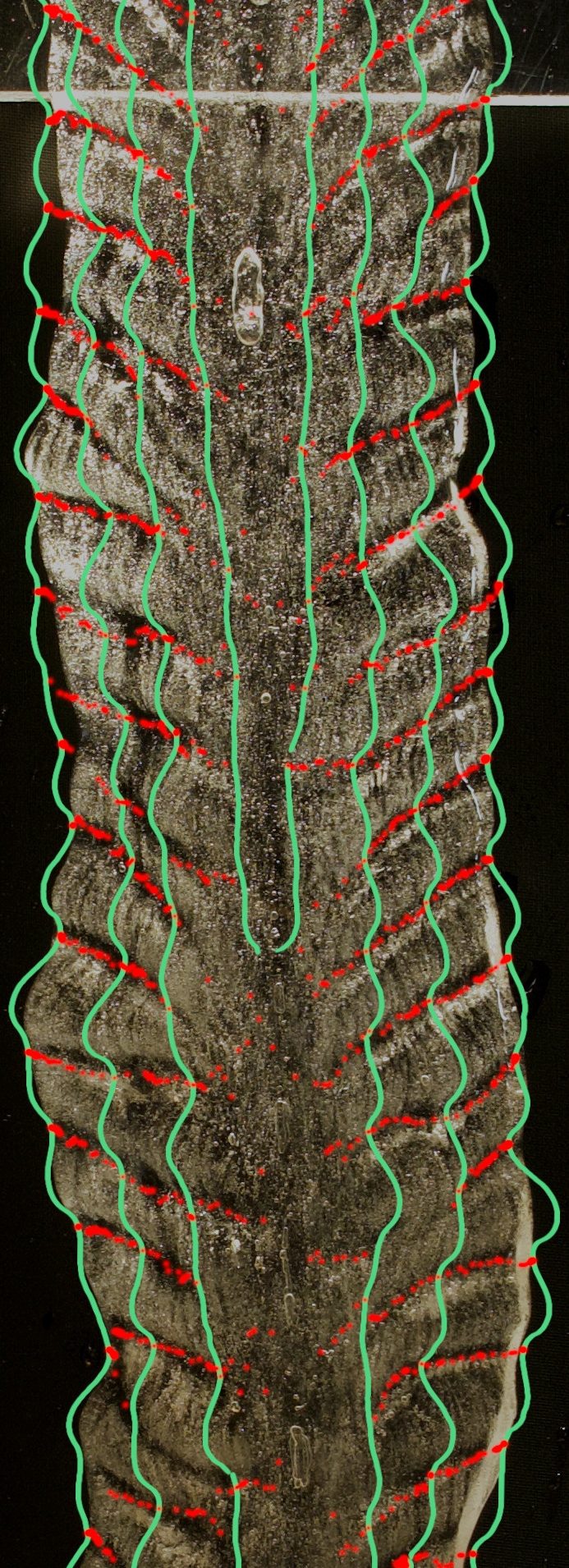}
    \end{center}
    \caption{An overlay of a cross sectional image with the edge-detected surface shape taken every 10 rotations shows that
    the bands of inclusions and crescent shapes closely aligns with the topography.
    The valleys between the ripples, highlighted in red, match up well with the
    lines of pure ice seen in the cross section.}
    \label{fig:icicle-overlay}
\end{figure}

Examining Fig.~\ref{fig:nacl-range}, it is obvious to the eye that the bands of
impurities inside the icicle are closely lined up with the exterior ripples.
The peaked shapes of the crescent structures match up with the shapes of the
ripple peaks, and the lines of pure ice align with the valleys between peaks.
Using time-lapse images taken during the icicle growth, we can follow the time
evolution of the topography and correlate it with a cross sectional image of the
final distribution of inclusions. Fig.~\ref{fig:icicle-overlay} illustrates this
close correspondence.  

It is frequently observed that ripples exhibit wavelength selection effects
during their evolution: ripples crowd one another causing ripples to disappear,
or ripples split, causing new ripples to appear between two existing ripples.
These processes can be seen in Fig.~\ref{fig:space-time-peaks}, which shows the
topography with peak tracking.  Similar ripple creation and destruction events
are traced out by bands of inclusions in cross sectional images, as in
Figs.~\ref{fig:nacl-range} and ~\ref{fig:section-measurement}.  These effects
are most pronounced at the highest concentrations, when the icicle is densely
filled with liquid inclusions.

In Secs.~\ref{sec:wavelength} and ~\ref{sec:peak-migration} below, we
quantitatively compare the chevron pattern of inclusions to the rippled
topography through their wavelengths and the migration of ripples up the icicle.

\subsection{Ripple wavelengths}
\label{sec:wavelength}

The ripple wavelengths were determined using a statistical approach, based on
real-space position measurements, both on edge detected topography and on cross
sections.
We measured the distance between adjacent features; peaks or valleys in edge
data, bands in cross sections.  The average of the distance measurements gives a
ripple wavelength consistent with previous measurements that
were made using a Fourier technique~\cite{chen2013njp}.
Maeno \etal{}~\cite{maeno1984b} also used the mean distance between peaks to measure wavelength.
This method allowed us to use data from a
larger fraction of the icicle, as well as giving a better statistical
understanding of the variability of the wavelength.

To obtain the ripple wavelength during growth, each icicle was imaged over 700
times, and both edges were detected, as described in Sec.~\ref{sec:methods}.
This results in tens of thousands of position measurements of ripple peaks and
valleys for each icicle. 
The wavelength was calculated independently for each edge (\ie{} the left and
right sides at each point in time), providing a wavelength at each point in time during
growth.
We found that the wavelength was constant throughout growth, so that the median value
of all these wavelengths could be taken to be the ripple wavelength for each
icicle during its growth.  

\begin{figure}
    \begin{center}
        \includegraphics[width=\columnwidth]{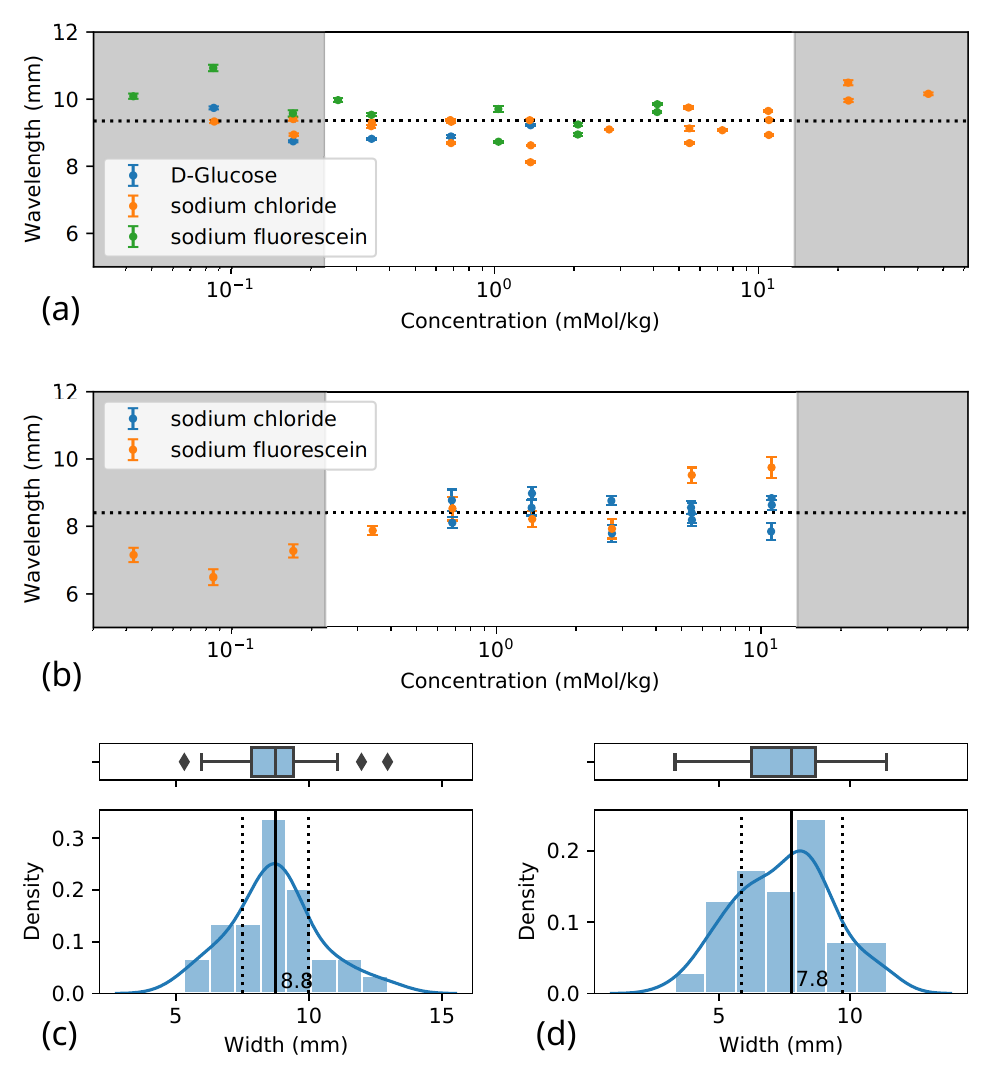}
    \end{center}
    \caption{Ripple wavelengths of icicles as measured from the (a) topography, and
    (b) cross-sections. Grey regions indicate concentrations where ripples are
    not ``well-behaved'' --- not forming at low concentrations, or exhibiting
    uncontrolled dynamics at higher concentrations, when the interior is fully
    saturated with inclusions.
    The error  bars indicate  the \textit{standard error of the mean} for each
    wavelength measurement.
    Sample distributions of individual ripple widths shown for (c) a single
    edge, and (d) one cross-section. Both distributions are for icicles with
    80~ppm~NaCl.}
\label{fig:ripple-length}
\end{figure}

We limited the edge sampling to a time and region where ripples are well-formed;
between 90 and 240~minutes into growth, and the top 20~cm of the icicle. 
Results of this analysis are shown in Fig.~\ref{fig:ripple-length}. Even within
these sampling limits, there remains a sampling bias caused by missing ripple
peaks/valleys, which skews the wavelength distribution towards longer ripples.
We use the mode of the
wavelength distribution for each edge (rather than the mean or median) to reduce the effects
of this sampling bias.  We find that the wavelength is 
independent of the concentration and  species of the
impurity used.
The average and variability of the wavelength of the icicles presented 
here are summarized in Tab.~\ref{tab:wavelengths}. Our results are consistent
with those of Maeno \etal{}~\cite{maeno1984b} and Chen
\etal{}~\cite{chen2013njp}, which are also included in the table.



\begin{figure*}
    \begin{center}
      \includegraphics[width=\textwidth]{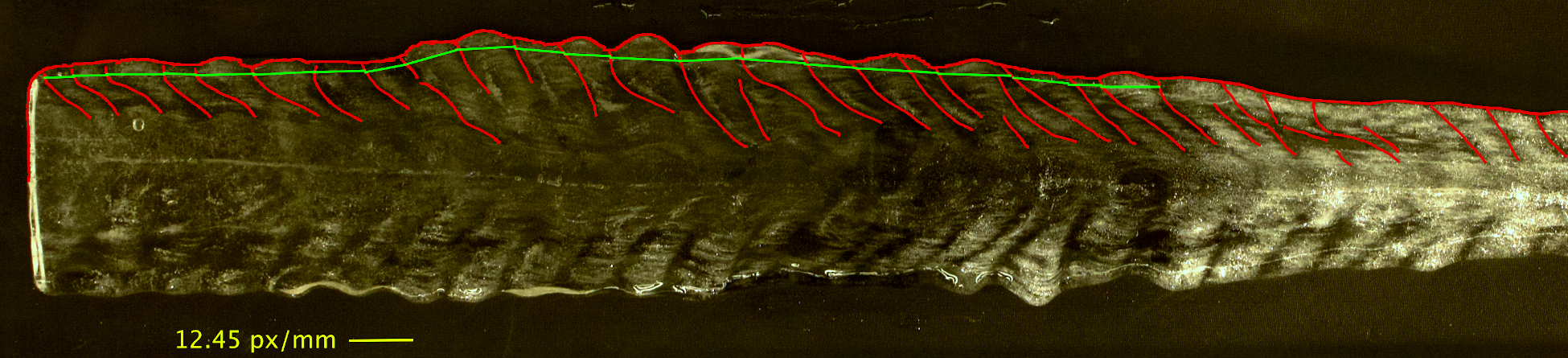}
    \end{center}
\caption{An example of measuring an icicle cross section. The icicle is illuminated
    with a unidirectional light source to the right. The pure ice that separates
    bands of impurities are highlighted with lines. The feature-widths (green
    lines) are measured by the distance between each band of inclusions. The
    distribution of widths can then be used to find the mean and variance of
    band-to-band distances.}
\label{fig:section-measurement}
\end{figure*}

An independent estimate of the chevron band wavelength can be obtained purely
from cross sectional images, albeit with smaller statistics. Longitudinal cross
sections were analyzed using ImageJ~\cite{schindelin2012Fiji}. The lines of pure
ice between bands of inclusions were traced by hand, and the distances between
adjacent lines were measured to obtain wavelengths. In practice, this rather manual procedure was
limited to a ``well behaved" concentration range where it was possible to easily distinguish and annotate the pure ice regions.
Below about 7~ppm~NaCl, coherent bands do not form, so
annotations are unreliable, while above 500~ppm~NaCl the ice is
saturated with inclusions, so distinct chevron bands can no longer be picked out by eye.
A sample measurement of the band wavelengths from cross sections in the well behaved region is shown in
Fig.~\ref{fig:section-measurement} and full results are shown in
Fig.~\ref{fig:ripple-length}.

We find the wavelength of the chevron band features to be $8.54 \pm 0.66$~mm 
for the 25 measured cross-sections in the well behaved concentration range. As
expected, the wavelength of the bands of inclusions (see
Tab.~\ref{tab:wavelengths}) agrees with the wavelength derived from the
topography.  Both agree with the earlier estimate of 9.0~mm by Maeno
\etal{}~\cite{maeno1994} and are slightly smaller than the Fourier value of
$10.4 \pm 0.8$~mm for the topography found by Chen \etal{}~\cite{chen2013njp}.


\begin{table}
    \caption{Summary of wavelength measurements for all icicles analyzed,
    including both data from edge detection and from annotated cross sections.
    Measurements from cross sections are more reliable at intermediate
    concentrations, where the bands are well behaved.
    }
    \label{tab:wavelengths}
    \begin{tabular}{r|cccc}
        Dataset & Number & Average & Standard &  70\% range \\
           & of icicles & (mm)   & dev. (mm)  & (mm)\\
        \hline
        Edges & 40 & 9.35 & 0.57 & 8.7 - 9.9 \\
          Sections:  (all)   &  30 &   8.30  &   0.68  &    7.78 - 9.06 \\
         (well behaved)  & 25  &  8.54   &  0.66  &  7.8 - 9.3 \\
         \hline
        Maeno \etal{}~\cite{maeno1994} &  19  & 9.0  &  & 7.0 - 10.0 \\
        Chen \etal{}~\cite{chen2013njp} &  67  & 10.4 & 0.8 & \\
    \end{tabular}
\end{table}

There is a slight trend toward longer wavelength at higher concentrations seen in
the cross section measurements, which was also observed by Chen
\etal{}~\cite{chen2013njp}, but it is not significant given our number of
measurements, and not reflected in the wavelengths measured from the topography
during growth. In the topography and in Chen's measurements, the
increase in wavelength is observed above the inclusion saturation  point of about
500~ppm~NaCl.

\subsection{Migration of the ripple peaks}
\label{sec:peak-migration}

In Fig.~\ref{fig:icicle-overlay}, it is clear that the bands of inclusions
closely coincide with the topography throughout the growth of the icicle, and
the upward tilt of the chevron bands tracks the motion of the rippled
topography. In order to quantify this effect, we compared the slope of the
chevron bands to the positions of the ripple peaks over time.

It is only possible to determine the angles made by chevron bands in cross
sections taken at intermediate concentrations, for which clear, well behaved
bands are traceable. Using the same traced lines of pure ice shown in an example
in Fig.~\ref{fig:section-measurement}, the typical angle of of the chevron bands
was found to be $58\pm 2.8^\circ$ to the icicle axis. Above 160~ppm~NaCl, we
observed much higher angle variability within each icicle. Below 160~ppm~NaCl,
the standard deviation was typically $10^\circ$, while above 160~ppm~NaCl it
could be as high as $35^\circ$. Thus, the speed of migration of the ripples becomes more variable well before the 
icicle becomes saturated with inclusions.

By tracking the peaks in the topography as in Fig.~\ref{fig:space-time-peaks},
we can extract the position of the peak relative to the icicle axis at each
point in time. Fitting a line to those measurements gives a typical angle of
$56\pm 2^\circ$ to the icicle axis, which is consistent with the angle made by
the chevron bands seen in the cross-section.

In icicles a higher concentrations that are saturated with liquid inclusions, we
no longer observe lines of pure ice in the cross sections. This occurs at a
concentration where Chen \etal{}~\cite{chen2013njp} found the direction of
ripple migration became more unpredictable and was often downwards. When we
track the peaks in the topography, we observe the same unpredictable ripple
migration at these concentrations. Ripple migration on saturated icicles is much
more dynamic than at lower concentration: peaks may split, consume neighbors, or
spontaneously form in valleys, as shown in Fig.~\ref{fig:space-time-peaks}(c).

We propose in Sec.~\ref{sec:discussion} that the direction of peak migration may
be understood to be a consequence of the distribution of surface liquid on
icicles.

\subsection{Crystal grain structure}
\label{sec:grain-structure}

\begin{figure}
    \begin{center}
        \includegraphics[width=\columnwidth]{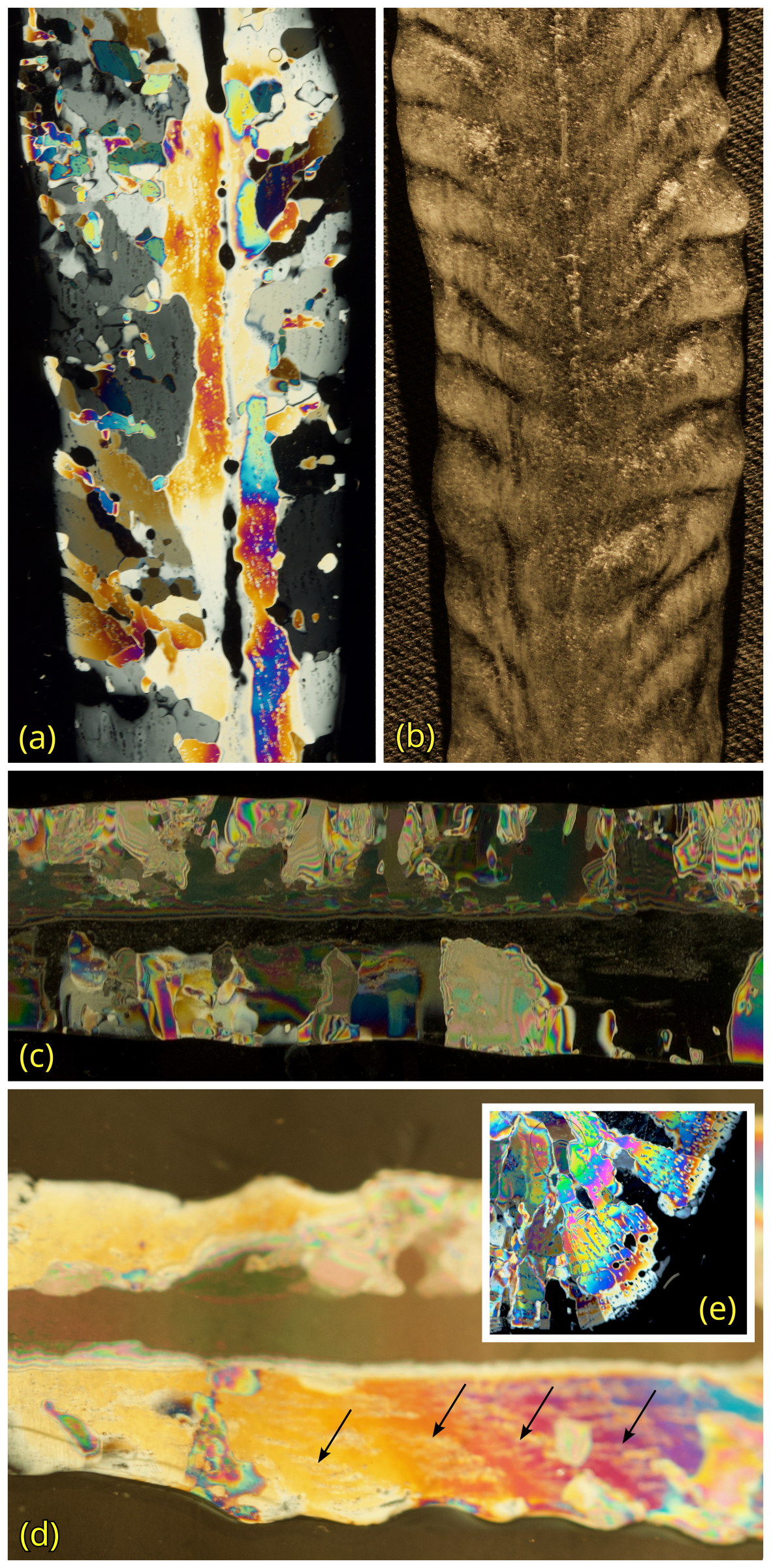}
    \end{center}
    \caption{Images of icicle cross sections between crossed polarizers. The upward angle of
    grains in a rippled icicle (80~ppm~NaCl) can be seen in (a), with its
    corresponding non-polarized view in (b). When an icicle is grown with a low
    impurity concentration (20~ppm~NaCl) no ripples form (c), and the grains do not exhibit
    an upward angle. The liquid inclusions are apparent in (d) (40~ppm~NaCl), which is
    also an example of a crystallite spanning multiple ripples. Arrows in (d) point to
    individual bands of inclusions seen within the single crystallite. The inset
    (e) shows grains with internal inclusions in an axial cross-section.}
    \label{fig:polarized}
\end{figure}

We used the same cross sectioning technique to examine the crystal structure of
icicle sections at various concentrations.
A selection of icicles viewed with cross-polarized filters are shown in
Fig.~\ref{fig:polarized}.
We found two consistent features in the grain structure of laboratory-grown icicles.
First, the core of the icicle near its axis was one long continuous crystallite.
There is generally a single crystal with its c-axis perpendicular to the icicle axis
down the core, where the central column of liquid
water freezes last. 
Secondly, outside the core, the crystallites near the root of the icicle are
smaller than those lower down the icicle. 
The smaller crystallites near the root could be caused by the more 
rapid flow conditions encouraging more nucleation sites.  

These two observations are consistent with what Laudise \etal{}\ reported for
natural icicles~\cite{laudise1979}, although their sample of natural icicles
showed much more variety, which is likely due to the uncontrolled growth
conditions. Laudise \etal{}\ also found that the slower growing 
c-axis was never parallel to the icicle axis and that it was often nearly
perpendicular. In general, outside the core, the crystal orientations of the
crystallites are rather random. 

A third feature that was less consistent for rippled icicles was that the
crystallites and grain boundaries outside the core were sometimes oriented with
an upward bias, in a pattern similar to the chevron bands of impurities. These
features are visible in Fig.~\ref{fig:polarized}(a,b). The grain boundaries were
somewhat more likely to coincide with the lines of \textit{pure ice}, with the
trapped impurities clearly visible within the crystallites, not at the grain
boundaries.  Non-rippled icicles do not exhibit an upward bias in their grain
shape. Instead, their grain boundaries are mostly perpendicular to the icicle
axis, as in Fig.~\ref{fig:polarized}(c).

The correlation between the ripples and the grain structure is not strong,
however. Often a crystallite will span multiple ripples, as in
Fig.~\ref{fig:polarized}(d), or the there will be many small grains within a
ripple, with only a barely discernible tendency towards a chevron-like pattern.

Our lab-grown icicles show the same ripples as natural icicles with the
advantage of being able to control the water source and growing conditions. We
have shown that the inclusions are primarily pockets of highly concentrated
liquid trapped inside of crystallites. These inclusions form a record of the
icicle growth --- both of the formation of the rippled morphology, and of the cyclic wetting/freezing process.

\section{Discussion}
\label{sec:discussion}

In this Section, we situate our observations in the context of previous work,
and discuss how the inclusion patterns in the interior of an icicle may be
related to previous observations of the flow pattern of liquid at the surface
during growth~\cite{ladan2021wetting}.  Finally, we consider how our
observations shed light on the overall problem of icicle morphology and the
mechanism of ripple formation.

Although foggy, rippled natural icicles are commonplace, there have been
relatively few published descriptions of their internal structure. 
Knight~\cite{knight1980} observed sponginess and what he called ``air bubbles'' in icicles that have
a ``wrinkled'' (\ie{}\ rippled) appearance, 
and observed rings of inclusions in \textit{transverse} cross sections.
Maeno \etal{}~\cite{maeno1984b,maeno1994} observed foggy features inside of icicles
coinciding with the ripples on the surface in a longitudinal cross section of an icicle collected in Sapporo
during the winter in 1984, and likewise attributed the features to air bubbles.
Both these studies
were made before the crucial connection between ripples and
impurities became clear in laboratory experiments~\cite{chen2013njp}.

Under the assumption that the inclusions were air bubbles, Knight thought that
the spongy ice was primarily due to draining of liquid, while Maeno \etal{}\
suggested that ``air bubbles'' were evidence of rapid crystal growth.
In addition, Maeno \etal{}\ assumed that the whole surface was wetted, so ripples
would have to be from more rapid cooling at the protuberances. 
This explanation led to models of icicle
morphology~\cite{short2006,makkonen1988Model} and linear stability
theories of icicle ripples~\cite{ogawa2002,ueno2003,ueno2004,ueno2007,ueno2010a,ueno2010b,ueno2011} 
which assumed that the
icicle is completely covered by a thin flowing film of supercooled water.
Subsequent experimental
work~\cite{ladan2021wetting} has shown that this is not the case and that the 
water actually flows over the surface in a complex pattern of transient rivulets.


Our observation that the inclusions are pockets of liquid indicates that a
different process is involved in ripply icicle formation. The foggy, spongey ice
is not due to draining of liquid 
{from spongy ice}, but actually the trapping of impure liquid.
Thus, the inclusions are \textit{not} evidence of rapid crystal growth, and may
not be used to infer rapid cooling at protuberances.

While dissolved solids remain trapped in the liquid inclusions, dissolved gasses
can escape to the atmosphere. For gasses to be trapped as bubbles, there would
indeed have to be very rapid cooling, or some process that entrains and traps
bubbles as the ice forms.  Such a trapping process does 
occur in the core of the icicle, where the water can remain liquid well away
from the tip, forming a long hollow region. This core water only freezes later
by a sort of pipe-filling process, called mode 3 by Maeno
\etal{}~\cite{maeno1994}. The shapes of bubbles formed during this process were
observed by Maeno \textit{et al.} However, we have observed using dye that
liquid inclusions may still form 
in the core. Most of the body of an icicle is formed from liquid flowing and
pooling on the outer surface of the icicle~\cite{ladan2021wetting}, where it is
presumably in sufficient contact with the surrounding air that dissolved gases
can escape before bubbles are formed.

We suggest that the location and trapping of impurities can best be understood
by taking into account the way liquid flows and is otherwise deposited on the
surface of an icicle.  Previous experiments~\cite{ladan2021wetting}, which
studied this process in detail, have shown that the flow proceeds by a series of
intermittent rivulets accompanied by areas of stagnant surface wetting, rather
than completely ensheathing the icicle.  The nature of these flows is strongly
dependent on the concentration of impurities, which change the wetting
properties of the surface.  It was observed directly that the liquid tends to
reside preferentially on the upper surface of a ripple
peak~\cite{ladan2021wetting}. The wetting of an icicle is highly intermittent,
which indicates a cyclic wetting/freezing process, as Knight suggested when he
observed rings of inclusions in his transverse cross-sections.

The layering of the inclusions into the crescent structures may be interpreted
as evidence of an iterated wetting/freezing process in the following way.  The
liquid that remains on the surface after a rivulet has passed is confined to a
small wetted region. When that liquid is first deposited, it has the low initial
concentration of the feed water, and pure ice can form unimpeded. Rejection of
the impurities from the ice surface into the small volume of surface water would
cause a rise in concentration in the liquid. Mushy
ice~\cite{worster1997Natural,worster1992Dynamics} can then start to form,
trapping some impurities as inclusions while the remaining impurities continue
to build up in the surface liquid. The next rivulet of liquid would then wash
away or dilute the concentrated liquid, leaving a deposit with a lower
concentration again. The freezing process can then repeat, forming a stack of
crescent-shaped mixed-phase ice and pure ice, which organize into bands of
impurities matching the rippled topography of the icicle.

The lines of pure ice that we observe in the cross sections are colocated with
the mostly unwetted regions on the ice surface, while the areas with the most
inclusions colocate with the areas that are wetted with surface liquid for the
largest fraction of the time. 
In addition, the surface wetting and rivulet
flow process is modified by the presence of impurities in such a way that even
for constant overall flow rate, much more surface water is present at higher
concentrations~\cite{ladan2021wetting}.

The rather low feed water concentration is far below the threshold for a mushy
layer to form.
However, because the liquid is confined on the surface, as it freezes, the
concentration will rise above the critical concentration, and thus a mushy layer
can form in a transient way.
{The confinement of impurities allows a high value of $k$,
the ratio of bulk ice to feed water concentration, of $0.27 \pm 0.05$, to be
reached, far larger than the segregation coefficient for planar ice growth $k_0$.
Our value of $k$ is similar to the lower limit value $k^*$ for spongy ice extrapolated to zero growth velocity~\cite{weeks1967Effective},
which would normally not be possible for such low concentrations. 

All of these ice deposition effects conspire to cause a concentration-dependent
rippling instability that corrugates the icicle topography into a pattern of
ripples with a near universal wavelength.  
Obviously, there is much that we do not understand about the exact nature of
this instability.

We also examined the crystal grain structure, because of anecdotal suggestions~\cite{walker1988SciAm} that 
grain orientations may be connected to ripple growth.
For example, Knight~\cite{knight1980} observed that a knife blade could be inserted into the
hollows of a ripply icicle, implying that the spongey parts were in the hollows. 
That simple observation led Knight to suggest that the preferred crystal
orientation would have the c-axis parallel to the icicle axis.
The liquid could then drain from between the basal sheets leading to trapped air bubbles.
When we tried to perform the same knife blade observation on finished icicles,
we found that, contrary to Knight, we could  not insert a knife blade into the
hollows, and that in fact the ice was mushiest near the peaks of the ripples.
Knight acknowledged that previous work~\cite{laudise1979} had found that the
c-axis was never parallel to the icicle axis. We hope that our observations help
to clarify the relationship between mushy regions, crystal orientations and
ripples.

The traces of the ripple pattern that we observe in the crystal grain structure
might also be interpreted as an effect of the surface
liquid dynamics. 
The observed 
incommensurability of the grain structure with the ripple topography
shows that the ripples and their migration are not caused by the crystal orientation
setting a preferred growth direction. 
Instead, we suggest that the growth of the ripples from the corresponding liquid deposits
restrict the growth of the grains.

The large uninterrupted crystal can form in the core of the icicle, because the
outside of the tip is always coated and its core is filled with liquid,
which permits a slow, steady growth.
The smaller grains near the top of the icicle are likely linked to the rapid change
in temperature and/or flow in that region, which causes more rapid freezing.  
These inlet effects would be peculiar to our icicle machine, which is why the natural
icicles observed by Laudise \etal{}~\cite{laudise1979} had much more variety in their grain structure.

The slight upward bias in the grain shape could be due to the liquid deposits being
preferentially found on the top side of ripple peaks. 
If a new crystal nucleates on the top of a ripple peak, it has access to a large
pool of water to steadily grow, whereas the ``dry'' area on the underside creates a barrier that limits grain growth
downwards.  Of course, all these effects are highly stochastic so that only a rather weak correlations between the ripples and crystal structure are observed.

\section{Conclusions}
\label{sec:conclusions}


We conducted the most complete analysis to date of the foggy patterns of
inclusions in the interior of icicles and their connection to ripples on the
surface. We show that the inclusions are not air bubbles, but are actually
pockets of water with high concentrations of impurities. The inclusions in the
ice are organized into chevron bands, with crescent shaped substructures that
match the growth and migration of ripples on the surface. The impurities trapped
in the ice do not coincide with crystal grain boundaries, but the orientation
and shape of the grains are slightly influenced by the growth mechanisms of the
ripples.

The location and pattern of trapped impurities records the complex dynamics of the liquid 
on the surface of an icicle during growth. The inclusions collect in layered
crescents, reflecting the intermittent wetting and freezing of liquid on the
surface. Higher concentrations of liquid are achieved through the stagnation of
liquid on the surface. This liquid tends to linger on the upper surface of ripples,
where the highest concentration of impurities is trapped in the ice.

While the exact mechanism underlying icicle ripple formation is still unclear, our results
give some insight into what a model of the instability must include. We propose that each step of the cyclic wetting/freezing process of ripple growth might be understood through mushy layer theory applied to a small, transient volume of liquid. This small scale process could be
probed experimentally in greater detail  by observing the freezing behaviour of discrete rivulets of
water flowing over ice with low concentrations of impurities. The macroscopic rippling instability could then be accounted for by examining the stability of an averaged model of the underlying small scale processes.

\bibliographystyle{unsrt}

\bibliography{inclusions}  

\end{document}